\documentstyle[12pt,epsfig]{article}
\def\baselinestretch{1.0}
\advance\textheight by \topskip
\begin{document}
\tolerance=100000
\thispagestyle{empty}
\setcounter{page}{0}
\def\lsim{\raisebox{-.1em}{$
\buildrel{\scriptscriptstyle <}\over{\scriptscriptstyle\sim}$}}
\def\gsim{\raisebox{-.1em}{$
\buildrel{\scriptscriptstyle >}\over{\scriptscriptstyle\sim}$}}
\def\preprint{{preprint}}
\begin{flushright}
{DESY-00-150}\\
{TUHEP-TH-00124}\\
{RAL-TR-2000-029}\\
{December 2000}\\
\end{flushright}

%%%\vspace*{\fill}

\begin{center}
{\Large \bf
Signatures of MSSM charged Higgs bosons \\[0.25 cm]
via chargino-neutralino decay channels at the LHC}\\[0.75 cm]
{\large Mike Bisset$^*$}\\[0.2 cm]
{\it Department of Physics,}\\
{\it Tsinghua University,}\\
{\it Beijing, P.R. China 100084.}
\\[0.7cm]
{\large Monoranjan Guchait$^*$}\\[0.2 cm]
{\it DESY Theorie, Notkestr. 85,}\\
{\it 22603 Hamburg, Germany.}
\\[0.7cm]
{\large Stefano Moretti$^{*,\dagger}$}\\[0.2 cm]
{\it Particle Physics Department,}\\
{\it Rutherford Appleton Laboratory,}\\
{\it Chilton, Didcot, Oxon OX11 0QX, UK.}
\\[0.45cm]
\end{center}
\vspace*{\fill}

\begin{abstract}
{\vskip0.25cm\noindent\small 
We assess the potential of detecting a charged Higgs boson of the 
MSSM at the LHC via its decays into a chargino and a neutralino.
We focus our attention on the region of parameter space with
$m_{H^{\pm}} > m_t$ and $3 \,\, \lsim \, \tan\!\beta \,\, \lsim \, 10$, 
where identification of the $H^{\pm}$ via other decay modes has
proven to be ineffective.  Searching for means to plug this
hole, we simulate the decays 
$H^{\pm} \rightarrow {\widetilde\chi}_1^{\pm}{\widetilde\chi}_1^0$
and
$H^{\pm} \rightarrow {\widetilde\chi}_1^{\pm}{\widetilde\chi}_2^0,
{\widetilde\chi}_1^{\pm}{\widetilde\chi}_3^0$ --- the former can yield
a single hard lepton (from the chargino decay) while the latter 
can yield three leptons (from the chargino and neutralino decays).
Coupled with the dominant top quark $+$ charged Higgs boson
production mode, the resulting signature is one or three
hard, isolated leptons, substantial missing transverse momentum 
and a reconstructed (via a 3-jet invariant mass)
top quark.  The single lepton channel is swamped by 
background processes; however, with suitable cuts, a trilepton
signal emerges.  While such a signal suffers from a low 
number of surviving events (after cuts) and is dependent on 
several MSSM input parameters (notably $M_{\scriptscriptstyle 2}$,
$\mu$, and slepton masses), it does fill at least some of the 
void left by previous investigations.
}
\end{abstract}
\hrule
\vskip0.35cm
\noindent
\hskip0.00cm{$^*$ E-mails: bisset@mail.tsinghua.edu.cn,
guchait@mail.desy.de, stefano.moretti@cern.ch.}\\
\hskip0.00cm{$^{\dagger}$ {Current address: Theory Division, CERN,
CH-1211 Geneva 23, Switzerland.}
\vspace*{\fill}
\newpage

\noindent

A pair of spin-$0$ charged Higgs bosons, $H^{\pm}$, 
arises in any Two Higgs Doublet Model (2HDM) alongside a trio of
neutral Higgs bosons --- the $CP=+1$ `light' $h$ and `heavy' $H$
(with $m_h < m_H$) and the $CP=-1$ `pseudoscalar' $A$.
The charged Higgs bosons have been at the focal point
of extensive studies since they have no Standard Model (SM)
counterpart, and thus could provide irrefutably clear evidence
of an extended Higgs sector and new physics beyond the SM.
On the other hand, it may be difficult to either 
distinguish one type of neutral 2HDM Higgs boson from 
the SM Higgs boson or observe more than one of the 
neutral species \cite{GunOrr,KunZwir,bbktd,ATLCMS}.
Embedding the 2HDM inside the attractive theoretical 
framework of supersymmetry (SUSY) yields 
(together with additional assumptions about minimal field
content and minimal number of new couplings) the 
Minimal Supersymmetric Standard Model (MSSM).
In the MSSM, {\em at tree-level}, the masses of all
Higgs bosons, along with their couplings to the SM fermions 
and gauge bosons, can be parametrized in terms of only two 
unknown input parameters, generally taken as the mass of one 
of the Higgs bosons 
(typically either $m_A$ or $m_{H^{\pm}}$) and the ratio of 
the vacuum expectation values (VEV's) of the up-type and 
down-type Higgs doublets (denoted by $\tan\!\beta$)~\cite{guide}.  

As is well-known \cite{toploop}, these tree-level relations can
receive substantial radiative corrections, most importantly 
enabling $m_h > M_Z$ (making the upper limit on $m_h$ in the MSSM 
${\sim}135\, \hbox{GeV}$~\cite{twoloop}).  However, the tree-level
relation between the masses of the charged Higgs bosons and $A$,
$m_{H^{\pm}}^2 = m_A^2 + M_{W^\pm}^2$, is almost invariably quite 
insensitive to such corrections~\cite{mhc-cor}.   
Properly taking into account the corrections to the light Higgs boson
mass, $m_{H^{\pm}}$ may still be related to $m_h$, and an indirect
lower bound of ${\sim}140\, \hbox{GeV}$~\cite{mhcbound} for
$\tan\!\beta \, \simeq \,3$-$4$ can be placed on the former due to
the thus-far null search for a Higgs boson at LEP2.  This bound
grows rapidly stronger as $\tan\!\beta$ is decreased while tapering
very gradually as $\tan\!\beta$ is increased  (staying in the
$110$-$125\, \hbox{GeV}$ interval for $\tan\!\beta \gsim 5$).
There are also other processes where charged Higgs bosons
(or $A^0$, to whose mass that of the $H^{\pm}$ is closely tied)
enter as virtual particles at the one-loop level.
These include neutral meson mixing ($K^0 \bar{K}^0$, $D^0 \bar{D}^0$,
or $B^0 \bar{B}^0$) \cite{FCNCpap,looppap},
$Z^0 \rightarrow b \bar{b}$ ($R_b$) \cite{looppap,EWprecis}, and
$b \rightarrow s \gamma$ decays \cite{FCNCpap,looppap,EWprecis,bsgam}.
The $b \rightarrow s \gamma$ decays are
generally thought to be the most constraining \cite{EWprecis}.   
Here restrictions on $m_{H^{\pm}}$ are linked to a number of MSSM
variables, notably including the masses of the lighter chargino
and the stops.  We reserve further comment on this somewhat complex
issue until after we present our results.
  
More direct limits on charged Higgs bosons come from
hadron collider searches\footnote{There are also direct
searches for charged Higgs boson pair production at LEP2;
however, bounds obtained in this way are relatively low,
$m_{H^{\pm}} > 77.5$-$78.6\, \hbox{GeV}$ \cite{LEPmhbound}.
At a future $500\, \hbox{GeV}$ $e^+e^-$ linear collider, this 
could increase to a potentially competitive ${\sim}210\, \hbox{GeV}$
\cite{Sopczak}.}~$\!$for lepton non-universality (excess taus) in 
top quark decays resulting from $t \rightarrow  bH^+$ followed by 
$H^+ \rightarrow \tau^+ {\nu}_{\tau}$ \cite{lightMHC}
and the charge-conjugate reactions\footnote{Hereafter, inclusion 
of charged-conjugate processes may be assumed unless explicitly 
excluded.} (excessive numbers of charmed final states in top decays 
resulting from $H^+ \rightarrow \bar{s}c$ may also be of use,
as well as $H^+ \rightarrow W^+b\bar{b}$~\cite{Wbb} for 
$\tan\!\beta \simeq 1$).
At the soon to commence Run II of the upgraded Fermilab Tevatron,
such channels will allow experimenters to scan the
MSSM parameter space {\em for large and small values of}
$\tan\!\beta$ roughly up to the kinematical limit of the
$t \rightarrow b H^+$ decay, $m_t - m_b$~\cite{Tevatron}.
The reason for the $\tan\!\beta$ dependence stems from the
couplings between a charged Higgs boson and top and bottom 
quarks, given by\footnote{Analogous formul\ae\ hold for the 
other two SM fermion generations.}
\begin{equation}
\sim\frac{g^2}{2M_{W^\pm}^2}H^+ \left( 
m_t \cot\!\beta \bar{t} b_L + m_b \tan\!\beta \bar{t} b_R
\right).
\label{Htbcoup}
\end{equation}
The square of this, which has a minimum at 
$\tan\!\beta \simeq 6\,$--$7$,
is proportional to the strength of either the 
$g\bar{b} \rightarrow \bar{t}H^+$ cross-section
or the $t \rightarrow bH^+$ decay width.  In the intermediate 
$\tan\!\beta$ region around this minimum, a Tevatron search for charged 
Higgs boson pair production, $q\bar{q}~{\rightarrow}~H^+H^-$, which 
mainly proceeds utilizing only gauge couplings, 
could be feasible \cite{Kosuke}
if the charged Higgs bosons are light enough --- certainly
$m_{H^\pm} \, \lsim \, m_t$.  As the mass of the charged Higgs boson
grows larger than $m_t$, simple phase space suppression will 
severely handicap pair production. 

Thus the likely legacy bequeathed to the Large Hadron Collider 
(LHC) will be the 
%task of searching for 
pursuit of
a {\em heavy} charged Higgs boson (with
$m_{H^\pm} \, \gsim \, m_t$).
At the LHC, the dominant production mechanism for heavy $H^{\pm}$
scalars 
is via the $2 \rightarrow 2$ reaction $gb \rightarrow tH^-$~\cite{gbtH} 
and the $2 \rightarrow 3$ reaction
$gg \rightarrow t\bar{b}H^-$~\cite{ggbtH}. 
Alternative production modes\footnote{The $bq \rightarrow bH^\pm q'$
mode of \cite{bq} can only be relevant for very high values
of $\tan\!\beta$.} are charged Higgs pair production, 
$gg,q\bar{q} \rightarrow H^+H^-$ \cite{qqggHH}, and
associated production,
$gg,q\bar{q} \rightarrow W^{\pm}H^{\mp}$ \cite{assocWH}.
The former suffers from a lack of phase space for $m_{H^{\pm}} > m_t$ 
and low quark parton luminosities ($q\bar{q}$) in the LHC protons or 
heavy propagator loop suppression ($gg$).  The latter suffers from a 
huge irreducible background induced by either $t\bar{t}$ and/or 
$W^+W^-$ production and decay, depending upon whether the charged Higgs
boson decays via $b\bar{t},W^-h$ or $\tau^- \bar{\nu}_\tau$ states
\cite{WHwh}.  For the preferred production mechanism, henceforth to be
collectively referred to as the top-$H^{\pm}$ production mode, the
$\tan\!\beta$ dependence follows from the square of Eqn.$\!$~(\ref{Htbcoup}), 
again making high and low $\tan\!\beta$ values more accessible.  
The connection between the $2\rightarrow 2$ and the $2\rightarrow 3$ 
reactions has been discussed numerous times before
\cite{tbdec1,gbNgg}; the former is obtained from the latter if one 
of the gluons splits into a $b \bar{b}$ pair, with one bottom
quark ({\it i.e.}, $b$- or $\bar{b}$-quark) then interacting with 
the remaining gluon while the other is assumed to act as a spectator.  
The appropriate procedure \cite{gbNgg,scott} for estimating the 
inclusive Higgs production cross-section is to combine the 
$2\rightarrow 3$ subprocess with the $2\rightarrow 2$
subprocess through the subtraction of a common logarithmic term
$\sim\alpha_s\log(Q^2/m_b^2)$.
%A possible simplified approach is to merely simulate either the
%$2 \rightarrow 2$ scattering or the $2 \rightarrow 3$ scattering
%and then scale the result by the just-noted correction factor.  
However, utilizing the $2 \rightarrow 2$ simulation for the 
kinematical event selection (and thereby tacitly assuming that the 
final state bottom quark manifest in the $2\rightarrow 3$ 
formulation is soft and thus untaggable) could lead to erroneous 
event-shape parameter distributions (transverse momentum, opening 
angles between jets, {\it etc.}) due to the possible presence of the 
extra, neglected $b$-jet.  Therefore, here we simulate both 
subprocesses:  $2 \rightarrow 2$ simulations are employed solely to
normalize the cross-sections, making use of this subtraction 
procedure, while $2\rightarrow 3$ simulations are used to implement 
our selection and acceptance cuts with the $b$-jet resulting from 
the bottom quark produced in the $2 \rightarrow 3$ reaction subject 
to the same acceptance, resolution and isolation criteria as any 
other jet in that event.

Several decay modes for heavy $H^{\pm}$ states 
(for branching ratio studies, see \cite{chbr}), have been analyzed
assuming the above top-$H^{\pm}$ production mechanism, including:
$H^- \rightarrow b \bar{t}$~\cite{tbdec1,tbdec}, {\em generally expected}
to be the dominant decay mode; $H^- \rightarrow s \bar{c}$
(in ATLAS study of \cite{tbdec}); $H^- \rightarrow W^- h$~\cite{Whdec};
and $H^- \rightarrow \tau^-\bar\nu_\tau$~\cite{taunu}.  
All these decay modes (that will hereafter be termed ``SM'' decays)
were simulated (including parton shower, hadronization and
detector effects) in either the ATLAS simulations of
\cite{tbdec,Whdec,taunu} and/or in \cite{LesHouches}, with the 
latter concluding that $H^{\pm}$ scalars with masses up to ${\sim}400\,
\hbox{GeV}$ can be discovered by the LHC, 
{\em but only if} $\,\tan\!\beta \, \lsim \, 3$ (which is in the
neighborhood of the indirect limit from LEP2) {\em or} 
$\tan\!\beta \, \gsim \, 25$.  The ATLAS studies roughly concur, 
adding that the 
$H^- \rightarrow \tau^-\bar\nu_\tau$ channel can push the high
$\tan\!\beta$ reach down below $20$ for 
$m_{H^{\pm}} \, \lsim \, 400\, \hbox{GeV}$, with a minimum at
$\tan\beta\approx 10$ when $m_{H^{\pm}}$ is close to $m_t$. 

The purpose of this paper is to assess the prospects for 
utilizing the thus-far neglected SUSY decay channels of 
the MSSM charged Higgs bosons to probe regions of the 
parameter space inaccessible at the LHC via the SM decay 
modes.  The charged Higgs boson can in fact decay 
predominantly via these SUSY modes, as illustrated in 
Fig.$\!$ 1 --- which also serves to highlight the potential 
significance of these SUSY channels in the large 
$m_{H^{\pm}}$ and intermediate $\tan\!\beta$ regions.  
In particular, we explore $H^\pm$ decays into
a chargino ($\widetilde\chi_i^\pm$) and a neutralino
($\widetilde\chi_j^0$); {\it i.e.},
$H^{\pm} \rightarrow {{\widetilde\chi}}^{\pm}_i {{\widetilde\chi}}^0_j$,
$i=1$ or $2$ and $j=1$, $2$, $3$, or $4$.  The lightest neutralino,
${{\widetilde\chi}}^0_1$, is assumed to be the stable lightest SUSY
particle
(LSP).  The decay width is given by \cite{GunHab}:
\begin{equation}\label{SUSYwidth}
\Gamma (H^{\pm} \rightarrow \widetilde\chi^{\pm}_i \widetilde\chi^0_j) =
\frac{g^2 \lambda^{1/2}[
(F_L^2+F^2_R)(m^2_{H^\pm} - m_{\widetilde\chi_i^\pm}^2
- m_{\widetilde\chi_j^0}^2) - 4  \epsilon F_L F_R m_{\widetilde\chi_i^\pm}
m_{\widetilde\chi_j^0}]}{16 \pi m^3_{H^\pm}},
\end{equation}
where $F_L$ and $F_R$ are as follows:
\begin{eqnarray}
F_L &=& \cos\beta[N_{j4}V_{i1} + {\scriptstyle \sqrt{\frac{1}{2}}}
(N_{j2}+N_{j1}\tan\theta_W)V_{i2}], \nonumber \\
F_R &=& \sin\beta[N_{j3}U_{i1} - {\scriptstyle \sqrt{\frac{1}{2}}}
(N_{j2}+N_{j1}\tan\theta_W)U_{i2}].
\label{eq:hdecay}
\end{eqnarray}
(For the $U$, $V$ and $N$ matrices, we have followed the notation of
\cite{GunHab}.)
Here, $m_{\widetilde\chi_i^\pm}$($m_{\widetilde\chi_j^0}$) are the masses
corresponding to the $\widetilde\chi_i^\pm(\widetilde\chi_j^0)$ states and
$\epsilon$ is the sign convention for the neutralino mass eigenstates.
Dependence on the additional MSSM input parameters
$M_{\scriptscriptstyle 1}$, $M_{\scriptscriptstyle 2}$, and $\mu$ 
enters from the gaugino/Higgsino mixing matrices via the $U$, $V$ and 
$N$.  $M_{\scriptscriptstyle 1}$ and $M_{\scriptscriptstyle 2}$ are the 
$U(1)_{\hbox{\smash{\lower 0.25ex \hbox{${\scriptstyle Y}$}}}}$
and
$SU(2)_{\hbox{\smash{\lower 0.25ex \hbox{${\scriptstyle L}$}}}}$
gaugino masses, respectively, and $\mu$ is the Higgsino mass 
parameter.  Grand Unified Theories (GUT's) predict gaugino unification
and $M_{\scriptscriptstyle 1} =
\frac{5}{3}\tan\!^2{\theta}_{\scriptscriptstyle W}
M_{\scriptscriptstyle 2}$,
as will be assumed in all numerical calculations.

Branching ratios (BR's) for the chargino-neutralino decay modes of 
the charged Higgs bosons are shown (along with the important SM 
decay BR's) {\it versus} $\tan\!\beta$ in Fig.$\!$ 2, choosing
$M_{\scriptscriptstyle 2} = 200\, \hbox{GeV}$ and 
$\mu = -120\, \hbox{GeV}$ as in Fig.$\!$ 1.  While this point is 
favorable for chargino-neutralino decays, it did not result from a 
exhaustive search for the optimal choice.  Three charged Higgs boson
masses are examined ($m_{H^{\pm}} = 200$, $300$, and $400\, \hbox{GeV}$).
For $m_{H^{\pm}} = 200\, \hbox{GeV}$, the only chargino-neutralino
decay channel open is ${{\widetilde\chi}}^{\pm}_1 {{\widetilde\chi}}^0_1$;
whereas, for $m_{H^{\pm}} = 300\, \hbox{GeV}$, the
${{\widetilde\chi}}^{\pm}_1 {{\widetilde\chi}}^0_2$ and
${{\widetilde\chi}}^{\pm}_1 {{\widetilde\chi}}^0_3$ channels are also 
accessible.  In fact, in this latter case the BR for 
$H^{\pm} \rightarrow {{\widetilde\chi}}^{\pm}_1 {{\widetilde\chi}}^0_2$ is
larger than that for 
$H^{\pm} \rightarrow {{\widetilde\chi}}^{\pm}_1 {{\widetilde\chi}}^0_1$
for $\tan\!\beta \, \gsim \, 2$.  
By the time $m_{H^{\pm}}$ reaches $400\, \hbox{GeV}$,
many chargino-neutralino decay channels have opened up and the 
situation becomes fairly complicated.
Decays to the heaviest charginos and neutralinos may well
generate cascade decays rather than (predominantly) decaying 
directly to the LSP.  This will introduce additional MSSM
parameter space dependence as well as complicate the analysis.
Further, as we shall see, for $H^{\pm}$ masses much beyond
this point, the lower top-$H^{\pm}$ production rate robs us 
(after the necessary cuts) of any signal events in the 
multilepton channels we will be investigating.
Therefore, there is considerable justification for
concentrating upon the $H^{\pm} \rightarrow
{{\widetilde\chi}}^{\pm}_1 {{\widetilde\chi}}^0_1,
{{\widetilde\chi}}^{\pm}_1 {{\widetilde\chi}}^0_2,
{{\widetilde\chi}}^{\pm}_1 {{\widetilde\chi}}^0_3$
channels in this exploratory study.
 
Note from the $m_{H^{\pm}} = 400\, \hbox{GeV}$ plot in 
Fig.$\!$ 2 that the sum of the various chargino-neutralino
modes (which is represented by the ``all SUSY'' curve) does
in fact dominate over the SM modes in the $\tan\!\beta$ range
of interest.  Note also that the combined BR's to all the 
sleptons remains under (and usually well-under) 2\% even 
though (with $m_{\tilde{\ell}}\, \simeq \, 150\, \hbox{GeV}$)
such decay modes are open.\footnote{Exceptions to this general rule 
are found if the stau masses are lowered to the edge of the 
LEP2 excluded region ($m_{\tilde{\tau}_1} \sim 90\, \hbox{GeV}$)
and $m_{H^{\pm}} \lsim 200\, \hbox{GeV}$ --- see 
Borzumati and Djouadi in \cite{chbr}.}
Charged Higgs boson decays into sfermions (squarks and sleptons), 
$H^- \rightarrow \tilde{q} {\tilde{q}}^{'\! *}, \tilde{\ell}
\tilde{\nu}_{\scriptscriptstyle \ell}^*$, are in fact heavily suppressed
compared to the chargino-neutralino decay modes 
(by $\sim M_{W^\pm}/m_{H^{\pm}}$); 
and, typically, these BR's do remain below the percent level.  

In Fig.$\!$ 2, the LEP2 bound on the mass of the chargino 
will exclude $\tan\!\beta$ values {\em above} a certain 
cut-off value.  The exact value of this LEP2 bound\footnote{This 
does drop considerably if the chargino becomes near degenerate 
with the LSP; however, nowhere in the regions of parameter space 
we will investigate does this occur.} depends slightly on the mass 
of the electron-flavor sneutrino.  
For a heavy $\tilde{\nu}_e$, the current LEP2 bound is 
$m_{\widetilde\chi^\pm_1} \, \gsim \, 103.2\, \hbox{GeV}$ \cite{W1LEP2}.
If $m_{\tilde{\nu}_e} \, \lsim \, 200\, \hbox{GeV}$, this 
bound is weakened by only a GeV or so; however, this small
change is enough to shift the upper limit on allowable
values for $\tan\!\beta$ from ${\sim}23$ to ${\sim}39$.  
Also, as noted earlier, low values of $\tan\!\beta$ are excluded 
by LEP2 searches for a (neutral) Higgs boson.  For this particular 
set of MSSM input parameters, we derive bounds of roughly
$\tan\!\beta \, > \, 2.8,2.4,2.2 (3.5,2.9,2.8)$ for 
$m_{H^{\pm}} = 200,300,400\, \hbox{GeV}$ based on the
current (potential) LEP SM Higgs boson mass bounds given in 
the first (second) paper of \cite{LEPmhbound}.    
Therefore, $3 < \tan\!\beta < 10$, the region where charged 
Higgs boson signatures from SM decays are virtually absent, and thus 
the region of primary interest to us in studying chargino-neutralino
decay modes, is not excluded.

In this work, to avoid the enormous QCD background,
only leptonic decays of the SUSY particles (sparticles) involved 
are considered.\footnote{Similar decays of
neutral MSSM Higgs bosons were studied in \cite{bbktd,bbkt}.}
Two specific signal types are analyzed:  events containing 
either one or three hard leptons\footnote{Hereafter, `leptons'
will refer to electrons and muons in general and irrespective
of sign; taus and neutrinos are not included.} accompanied by 
missing transverse momentum, ${p}^{\mathrm{miss}}_T$, and a 
reconstructed top (meaning a $t$- or $\bar{t}$-quark).  
The top resonance is identified through the invariant mass of the 
three (at least at the parton level)
jets resulting from its hadronic decay.  
(Here, we consider the rate for mis-identifying tops as very 
low and disregard any backgrounds that could arise from such
mis-identification.)  
Tops decaying leptonically into a $b$-jet along with a charged 
lepton and a neutrino are not deemed to be part of the signal 
processes, 
but do play a r\^ole in potentially serious backgrounds and 
are included in all simulations (as are hadronically-decaying 
charginos and neutralinos).  
The leptons result from the following three-body
chargino and neutralino decays:
%%%chargino and neutralino decays ($j=2,3$),  
\begin{eqnarray}
\;\;\;\;\;\;\;\;\;
{{\widetilde\chi}}^{\pm}_1 \rightarrow 
{{\widetilde\chi}}^0_1 {\ell}^{\pm} 
\stackrel{{\scriptscriptstyle (-)}}{\nu_{\scriptscriptstyle \ell}} 
\;\;\; \hbox{and} \;\;\;
{{\widetilde\chi}}^0_j \rightarrow
{{\widetilde\chi}}^0_1 {\ell}^+ {\ell}^- \; (j=2,3) \; .
\label{inolep}
\end{eqnarray}
Note that the charginos and neutralinos decay directly to the LSP.
Appropriate BR's for these decays are incorporated if necessary,
but these are usually the only available decay modes.  If the only
virtual intermediate particles involved in these decays are the 
$W^{\pm}$ and the $Z^0$, then the leptonic BR's are the well-known
leptonic BR's of the intermediate vector bosons ($0.212$ and $0.067$, 
respectively).
However, if sleptons are relatively light, they can also 
mediate these decays and significantly enhance the leptonic 
BR's \cite{BaerTata} (especially those of the neutralinos),
and thus also the rates for our prospective signals.  
Such light sleptons
(with $m^{\rm{soft}}_{\tilde{\ell}} \sim 150\, \hbox{GeV}$)
are not excluded experimentally and would not be out of place in 
the light MSSM sparticle spectrum under consideration.   

In choosing the amount of missing transverse momentum required by
the cuts, some knowledge of the chargino and neutralino mass spectrum 
is presumed to be available from independent measurements \cite{LHCino}.
However, the charged Higgs boson mass is treated as a completely
unknown parameter.  Furthermore, due to the multiple particles leaving 
the detector unobserved, reconstruction of the Higgs boson and sparticle
masses involved in the signal decays is not possible.  Rather we
here content ourselves with looking for excesses in the specified modes
above the SM expectations.

The analysis presented here is confined to the parton level 
only\footnote{Although the signal processes have already been 
incorporated into the event generators HERWIG \cite{herwig} and ISAJET
\cite{isajet}, the incorporation of several important background 
processes (see below) is still in progress \cite{preparation}.}:
jets are identified with the partons from which they
originate and jet selection criteria are applied directly to the partons.
Typical detector resolutions (and range limitations) are included:
the transverse momenta of all visible particles in the final state have
been smeared according to a Gaussian distribution,
with $(\sigma(p_T)/p_T)^2 = (0.6/\sqrt{p_T})^2 + (0.04)^2$ for all
jets and
$(\sigma(p_T)/p_T)^2 = (0.12/\sqrt{p_T})^2 + (0.01)^2$ for the leptons.
The missing transverse momentum has been evaluated from
the vector sum of the jet and lepton transverse momenta after 
resolution smearing.  
For reference,
the CTEQ4L \cite{cteq4} structure function set is used, with
the factorization scale set to $Q = m_t + m_{H^{\pm}}$ for the Higgs 
boson processes and $Q = m_t$ for all others.  
Aside from using running quark masses and loop-corrected Higgs boson
masses, other higher order corrections to the tree-level top-$H^{\pm}$ 
production \cite{prodcorr} and hadronic $H^+ \rightarrow t \bar{b}$
decay \cite{decaycorr} (which competes with our preferred SUSY 
decay modes) are not taken into account.  The literature indicates
that these corrections will not change the results much, though with
small signals they should be kept in mind. 

Following values given in \cite{ATLAS-TDR}, we assume a
single $b$-tagging efficiency of $\epsilon_b=0.5$ and a
mis-tagging rate of $\epsilon_{mis}=0.02$ (though we
note the latter value may be too low since the study in
\cite{ATLAS-TDR} did not include $c$-quarks).
However, the $\bar{b}$-quark manifest in
$gg \rightarrow \bar{b}tH^-$ is often expected to be soft
and/or near the beam pipe.  Thus, $\epsilon_b=0.5$ is
probably a serious over-estimation in this case. 
It is inappropriate to graft a serious $b$-tagging study 
onto this parton-level analysis.  A more thorough
treatment will presented in the upcoming event
generator analysis \cite{preparation}.  As a simple
approximation we adopt an on-off switch:  if a 
$b$-jet (recall this is equivalent to a bottom quark) 
has a $p_T$ above a certain specified value
and an $|\eta|$ below another specified value, we assign
a $b$-tagging efficiency of $\epsilon_b=0.5$ to it; 
otherwise, we set $\epsilon_b=0$ for that soft and/or
too close to the beam pipe $b$-jet.  Those $b$-jets 
stemming from top decays are expected to almost always
pass this test, so we assume $\epsilon_b=0.5$ for all
such $b$-jets.  Now, fortuitously, it so happens that, 
for the particular case of one {\it versus} two $b$-jets 
and $\epsilon_b=0.5$, it does not matter how often there 
is one $b$-jet {\it versus} how often there are two $b$-jets 
fulfilling such criteria, since 
$\epsilon_b = 2\epsilon_b(1 - \epsilon_b) = 0.5$,
and so the overall $b$-tagging efficiency for the event 
will remain $\frac{1}{2}$.  
This is the case for the signals we are searching for as 
well as for all the backgrounds we will discuss.

We require one $b$-tagged jet in each event.
This in fact reduces both the event rates {\em and}
the signal to background ratios for the backgrounds
we will consider explicitly.  However, it also aids in
triggering and the suppression of incidental QCD 
backgrounds which we do not attempt to calculate.
In addition, we do veto events with more than one
$b$-tagged jet.  This does help in background
reduction for all the backgrounds we consider
as well as eliminating other bottom-rich
event-types (such as $gb \rightarrow t\bar{t}b$, 
$g\bar{b} \rightarrow t\bar{t}\bar{b}$,
$gg \rightarrow t\bar{t}b\bar{b}$, {\it etc.}).  
We do not identify individual 
$b$-quarks as tagged or untagged, rather we
multiply the event by a factor consistent with
the values given above.  Again, a more technical
treatment is inconsistent with this parton-level 
analysis and will come with the event generator  
studies.

\vskip0.25cm\noindent 
\underbar{\bf The one-lepton signature: $\ell^\pm 
+ {p}^{\mathrm{miss}}_T + t$}
\vskip0.25cm\noindent
If the charged Higgs boson is just above the top threshold,
$m_{H^{\pm}} \, \simeq \, 200\, \hbox{GeV}$, then it is quite likely 
that the only chargino-neutralino decay channel open will be 
$H^- \rightarrow \widetilde\chi_1^- \widetilde\chi_1^0$, as is the
case in Fig.$\!$ 2.  Schematically, the one lepton plus top signal 
would result from the reaction chain
\begin{equation}
gg \rightarrow \bar b t H^-, \quad
t \rightarrow b q \bar q', \quad
H^- \rightarrow \widetilde\chi^-_1 \widetilde\chi_1^0, \quad
\widetilde\chi^-_1 \rightarrow \widetilde\chi_1^0 \ell^- 
\bar{\nu}_{\scriptscriptstyle \ell} \; ,
\label{onelepton}
\end{equation}
($\ell=e,\mu$ and $q=d,u,s,c$).  
The hard lepton is derived from the decay of the chargino.
The leptonic BR of the chargino is generally not as strongly 
affected by a light slepton as are those of the non-LSP neutralinos,
though modest enhancement over the expectations from $W^{\pm}$-mediated
decays are possible.  

Fig.$\!$ 3 gives contour plots 
for BR($H^{\pm} \rightarrow \widetilde\chi^{\pm}_1 \widetilde\chi_1^0$)
with $m_{H^{\pm}} = 200\, \hbox{GeV}$, $\tan\!\beta = 4$, and
varying $M_{\scriptscriptstyle 2}$ and $\mu$ --- as noted
earlier, these are the main other MSSM parameters to which the 
chargino and neutralino properties are sensitive.  
The region of parameter space excluded by the LEP2 
bound on the chargino mass is indicated by the dotted curves
which are, from bottom to top, contours for 
$m_{\widetilde\chi^\pm_1} = 100$, $105$, and $110\,\hbox{GeV}$.
Note that the sensitivity to the exact bound here is 
much less than that of the $\tan\!\beta$ variable at
the upper end of its range (see discussion of Fig.$\!$ 2).
BR's for the desired charged Higgs boson decay channel in excess 
of 60\% are possible in unexcluded MSSM parameter space even with 
this relatively low charged Higgs boson mass.  
Guided by the study of this BR, we have selected the following 
point in the MSSM parameter space for detailed simulations
of the phenomenology of the one-lepton signature:
\begin{eqnarray}
M_{\scriptscriptstyle 2} =115\, \hbox{GeV}, \quad
\mu = -200\, \hbox{GeV}, \quad
\tan\!\beta=4\, , \quad 
m_{H^\pm} = 200\, \hbox{GeV} \; .
\label{point1l}
\end{eqnarray}
At this point, relevant masses and BR's are:
\begin{eqnarray}
m_{\widetilde\chi^\pm_{1,2}} = 112.61,~231.73\, \hbox{GeV},\quad
& & m_{\widetilde\chi^0_{1-4}}
=59.86,~111.16,~219.43,~221.63\, \hbox{GeV}, 
\nonumber \\
{\mathrm{BR}}(H^- \rightarrow \widetilde\chi_1^- \widetilde\chi_1^0) =
0.56\, , \quad\quad & &
{\mathrm{BR}}(H^- \rightarrow b\bar t) = 0.36 \;, 
\nonumber \\
{\mathrm{BR}}(\widetilde\chi_1^- \rightarrow \widetilde\chi_1^0 \ell^-
\bar{\nu}_{\scriptscriptstyle \ell}) = 0.28 \, \quad\quad & &
\!\!\!\!\!\!\!\!\!\! \!\!\!\!\!\!\!\!\!\!
(\hbox{for} \; m^{\mathrm{soft}}_{\widetilde\ell} = 150\, \hbox{GeV}) \;. 
\label{p1lchar}
\end{eqnarray}

SM backgrounds to such events come from top pair production 
and single top production:
\begin{eqnarray}
gg,q\bar q \rightarrow t \bar{t} \; ,  \quad
& t \rightarrow b q \bar q'\; , & \quad
\bar t \rightarrow \bar b \ell^- 
\bar{\nu}_{\scriptscriptstyle \ell} \; ,
\label{tt}
\\
gg,q\bar q \rightarrow t \bar{b} W^- \; ,  \quad
& t \rightarrow b q \bar q' \; , & \quad
W^- \rightarrow \ell^- 
\bar{\nu}_{\scriptscriptstyle \ell} \; ,
\label{onet}
\end{eqnarray}
where the initial $\bar b W^-$ pair in (\ref{onet}) does not 
come from an on-shell top decay.  At the LHC, approximately 
0.1 billion $t\bar{t}$ events will be produced for every 
$100\, \hbox{fb}^{-1}$ of integrated luminosity; whereas, 
the corresponding number of top-$H^{\pm}$ events is only 
several thousand --- note that 
(\ref{onet}) is suppressed relative to (\ref{tt}) by $\sim
\alpha_{\mathrm{em}}$,
meaning that rates for both backgrounds are larger than that
of the would-be signal.  
Finally, the bottom-top decay of the charged Higgs boson may 
yet have an appreciable BR even when chargino-neutralino decay 
modes are very important.  
Such `$H^- \rightarrow b\bar t\,$' events; {\it i.e.},
\begin{eqnarray}
& gg \rightarrow \bar b t H^- \; , & \quad
t \rightarrow b q \bar q'\; , \quad
H^- \rightarrow b\bar t \; ,\quad
\bar t \rightarrow \bar b \ell^- 
\bar{\nu}_{\scriptscriptstyle \ell} 
\nonumber \\
\hbox{or} \;\;\;\;\;\;
& gg \rightarrow \bar b t H^- \; , & \quad
t \rightarrow b \ell^+ {\nu}_{\scriptscriptstyle \ell} \; , \quad
H^- \rightarrow b\bar t \; ,\quad
\bar t \rightarrow \bar{b} \bar{q} q' \; ,
\label{tb}
\end{eqnarray}
might also pass our signal cuts, though these are not designed to 
optimize the selection of $H^- \rightarrow b\bar t$ events which, 
for instance, have four $b$-jets manifest in the decay chains of 
(\ref{tb}), whereas only one tagged $b$-jet is permitted by our
selection criteria.
To cut out additional QCD backgrounds, such as the radiation of 
hard gluons from the afore-mentioned SM backgrounds
(or MSSM gluino pair production followed by cascade decays), we 
will put a 4-jet cap\footnote{Of course, the jet number in an event
can be affected, for example, by the merging of showers from different
initial partons or hadrons lying too close to the beam pipe.  A full
event-generator analysis should yield a more accurate estimate of the
fraction of the time an expected jet is not seen than the present
parton-level analysis; so herein we will neglect such additional 
backgrounds.} on the number of jets we allow in any event.
Adding the four $b$-jets above and two distinct (neglecting the 
rare case of jet mis-identification) untagged jets we will
require to reconstruct a hadronically-decaying $W^{\pm}$
yields six jets, meaning most $H^- \rightarrow b\bar t$ events 
will also be lost to the 4-jet cut.\footnote{Because of this, 
associated production, $gg,q\bar{q} \rightarrow W^+H^-$, with the 
$W^+$ providing the hard lepton and the top coming from 
$H^- \rightarrow b\bar t$, might be a mimic of comparable size to 
(\ref{onelepton}), our designated reaction chain,  
(even though the $W^{\pm}H^{\mp}$ production cross-section at 
this MSSM point is down by roughly an order of magnitude from
top-$H^{\pm}$ production) if a higher fraction of such events 
survive the cuts.  However, since we will show explicitly that 
the reaction chains in (\ref{tb}) have a negligible effect, 
it is clear that this alternative reaction (or the even more 
suppressed charged Higgs pair production reaction chains) will 
also be unimportant.}  Combined with the compulsory single
tagged $b$-jet, this also implies that the surviving 
$H^- \rightarrow b\bar t$ events will have at least one and
at most two $b$-jets passing our on-off switch criteria, and so 
the $b$-tagging efficiency factor for these surviving events will 
again be $\frac{1}{2}$.
%
%Depending on your point of view, these events 
%could be considered as background to
%$H^- \rightarrow \widetilde\chi^-_1 \widetilde\chi_1^0$
%or as additional signal for a charged Higgs boson
%above the expected SM background.
   
In our simulation, we have adopted the following acceptance and 
selection cuts:
\begin{enumerate}
\item Jets and leptons are retained if they satisfy the
following requirements:
$p_T^{\ell,j}>25\, \hbox{GeV}$, $|\eta_{\ell,j}|<2$ and
$\Delta R_{\ell,j/j,j} =
\sqrt{\Delta \eta_{\ell,j/j,j} ^2 + \Delta \phi_{\ell,j/j,j}^2} > 0.4$,
where $j$ represents both $b$-tagged and untagged jets.
To pass this first cut, an event must have one and only one 
lepton that satisfies the above criteria and no more than 
four jets (irrespective of whether or not the jets are $b$-tagged)
fulfilling the requirements.

\item 
We require that ${p}^{\mathrm{miss}}_T > 80\, \hbox{GeV}$.

\item 
We demand that two untagged jets reproduce an
invariant mass around $M_{W^\pm}$:
\newline
$|M_{q\bar q'} - M_{W^\pm}|< 10\, \hbox{GeV}$.
Therefore, to take into account the possibility of 
mis-tagging a jet as a $b$-jet, we will multiply 
the $b$-tagging factor to be applied at the end of the 
series of cuts by
$(1 - 2\epsilon_{mis} + \epsilon_{mis}^2)= 0.96\,$.

\item 
We combine these two light-quark-jets with a $b$-jet
and demand that at least one such resulting 3-jet invariant 
mass be in the vicinity of $m_t$:
$|M_{b q \bar q'} - m_t| < 25\, \hbox{GeV}$.

\item 
Recognizing the difference in the number and type of
particles leaving the detector undetected in the signal
events (which have two LSP's as well as neutrinos) and 
background events (which have only neutrinos -- no LSP's), 
we construct a variable to exploit this distinction, 
demanding that
$\frac{|{p}^{\mathrm{miss}}_T-p_T^\ell|}
      {|{p}^{\mathrm{miss}}_T+p_T^\ell|} > 0.2$.

\item 
Finally, we apply a veto on a leptonically decaying 
$t$- or $\bar{t}$-quark.
If more than three jets (that is, jets in addition to the three jets
already assigned to a hadronically-decaying top in 3.$\!$ and 4.$\!$)
are present, then an invariant mass denoted by 
$M_{b \ell \nu_{\scriptscriptstyle \ell}}$
is formed from each extraneous jet's four-momentum and those of the
hard lepton and the missing momentum.  The missing momentum is
assumed to be solely due to a massless neutrino whose longitudinal
momentum is reconstructed following the technique outlined in the 
first paper of \cite{tbdec}.  This assignment is quite reasonable 
for SM double- and single-top events,\footnote{Since neutrinos that may
produced in decays of $B$-mesons or further on down the decay chains
inside the tagged and untagged $b$-jets are generally fairly soft.} so that
the former can be eliminated with a cut of 
$|M_{b \ell \nu_{\scriptscriptstyle \ell}} - m_t| > 25\, \hbox{GeV}$.
%%Note also that for the SM backgrounds (\ref{tt}) and (\ref{onet})
%%the extra jet is expected to be a (possible untagged) $b$-jet.
The assignment is of course not at all reasonable for decays of the
charged Higgs boson, and simulations confirm that a far smaller 
fraction of these events are lost in comparison to the
percentage of single- and double-top events weeded out.
\end{enumerate}

Results for this series of cuts are given in Tab.$\!$~\ref{tab_1l}.
With an integrated luminosity of $100\, \hbox{fb}^{-1}$, about
265 signal events per year of run time survive.  However,
approximately 50,000 background events also survive.
Even if one considers more favorable points in the MSSM parameter space,
it is very difficult to enhance the signal cross-section significantly.
Thus, if this one-lepton channel is to be useful in searching for 
charged Higgs boson at the LHC, far better cuts than those designed
here will need to be devised.

\begin{table}[!t]
\begin{center}
\begin{tabular}{|c|c|c|c|c|}
\hline
&&&&\\[-4mm]
& $t \bar t$
& $t \bar bW^-$
& $H^- \rightarrow b\bar t$ 
& $H^- \rightarrow \widetilde\chi^-_1 \widetilde\chi_1^0$\\
\hline
&&&&\\[-4mm]
No cuts
& 550. & 71.  & .20    & .30    \\
&      &      &        &        \\[-4mm]
\hline
&      &      &        &        \\[-4mm]
1 lepton with  
$p_T^\ell >25\, \hbox{GeV}$,
&      &      &        &        \\
$|\eta_\ell| < 2$,  $\Delta R_{\ell,j}>0.4 $
& 353. & 36.  & .15    & .120   \\
&      &      &        &        \\[-4mm]
\hline
&      &      &        &        \\[-4mm]
$\le 4$ jets with $p_T^j >25\, \hbox{GeV}$,
&      &      &        &        \\
$|\eta_j|<2$, $\Delta R_{j,j}>0.4 $
& 153. & 16.  & .033   & .042   \\
&      &      &        &        \\[-4mm]
\hline
&      &      &        &        \\[-4mm]
${p}^{\mathrm{miss}}_T > 80\, \hbox{GeV}$
& 28.  & 3.53 & .0049  & .025   \\
&      &      &        &        \\[-4mm]
\hline
&      &      &        &        \\[-4mm]
$|M_{q\bar q'}-M_{W^\pm}|<10\, \hbox{GeV}$
& 27.  & 3.38 & .0040  & .019   \\
&      &      &        &        \\[-4mm]
\hline
&      &      &        &        \\[-4mm]
$|M_{b q \bar q'} - m_t| < 25\, \hbox{GeV}$
&  25. & 2.93 & .0037  & .018   \\
&      &      &        &        \\[-4mm]
\hline
&      &      &        &        \\[-4mm]
$\frac{|{p}^{\mathrm{miss}}_T-p_T^\ell|}  
           {|{p}^{\mathrm{miss}}_T+p_T^\ell|} > 0.2$
& 18.  & 2.65 & .0029  & .017   \\
&      &      &        &        \\[-4mm]
\hline
&      &      &        &        \\[-4mm]   
$|M_{b \ell \nu_{\scriptscriptstyle \ell}} - m_t| > 25\, \hbox{GeV}$
& 1.99 & 1.49 & .0005  & .014   \\
\hline
\end{tabular}
\end{center}
\caption{\small Production and decay rates (in picobarns) for 
one-lepton signal and backgrounds, after the implementation of 
successive cuts.
Rates already include the Higgs boson decay BR's, whereas the
common $b$-tagging$\times$mis-tagging factor of
$\frac{1}{2} \! \times \! 0.96$ is not included.
Also omitted are the BR for one top to decay hadronically ($0.699$) 
and the leptonic BR's, $0.212$ for the top decays and 
${\mathrm{BR}}(\widetilde\chi_1^- \rightarrow \widetilde\chi_1^0 \ell^-
\bar{\nu}_{\scriptscriptstyle \ell})$ for the signal, as well as the
consequent combinatorial factor of 2.  
}
\label{tab_1l}
\end{table}
\newpage
\vskip0.25cm\noindent
\underbar{\bf
The three-lepton signature: $\ell^\pm\ell^-\ell^+ + {p}^{\mathrm{miss}}_T
+ t$}
\vskip0.25cm\noindent

For larger charged Higgs boson masses, other chargino-neutralino
decay modes besides $H^- \rightarrow \widetilde\chi_1^-
\widetilde\chi_1^0$
may well be open and have sizable BR's
(as shown in Fig.$\!$ 2).  The heavier neutralinos may then decay
leptonically (\ref{inolep}) and, together with the chargino, produce
three hard leptons (as first discussed in \cite{thesis}).  
Recall that in our notation `$\ell$' stands for either electrons
or muons.  For the signal, two leptons with opposite signs must be of 
the same flavor; the third lepton may also be of the same flavor or of
the other flavor.
The expected reaction chain for the signal is 
\begin{equation}
gg\rightarrow \bar{b} t H^- \; , \quad
t \rightarrow b q \bar{q}' \; , \quad
H^- \rightarrow \widetilde\chi^-_1 \widetilde\chi_{2,3}^0 \; , \quad
\widetilde\chi^-_1 \rightarrow \widetilde\chi_1^0 \ell^- 
\bar{\nu}_{\scriptscriptstyle \ell}
\;  ,\quad
\widetilde\chi^0_{2,3} \rightarrow \widetilde\chi_1^0 \ell^- \ell^+ \; .
\label{threelepton}
\end{equation} 
Fig.$\!$ 4 gives contour plots
for BR($H^{\pm} \rightarrow \widetilde\chi^{\pm}_1 \widetilde\chi_2^0$)
(on top) and
BR($H^{\pm} \rightarrow \widetilde\chi^{\pm}_1 \widetilde\chi_3^0$)
(on bottom),
with $m_{H^{\pm}} = 300\, \hbox{GeV}$, $\tan\!\beta = 4$, and
again varying $M_{\scriptscriptstyle 2}$ and $\mu$.  Limits of 
the regions excluded by LEP2 are again marked by dotted lines.  
BR's above 20\% are found at viable points in the parameter space.
(The diagonal discontinuities seen in the upper right corners of the 
$\mu < 0$ plots and the upper left corners of the $\mu > 0$ plots
are due to a `level-crossing' where the masses of $\widetilde\chi^0_2$ 
and $\widetilde\chi^0_3$ become degenerate --- thus the identities of
these 
two neutralinos are effectively interchanged as one of these
diagonal lines is crossed.)
From the study of these BR's the following point in the MSSM
parameter space was chosen for the simulation study:
\begin{eqnarray}
M_{\scriptscriptstyle 2} =200\, \hbox{GeV}, \quad
\mu = -120\, \hbox{GeV}, \quad
\tan\!\beta=4, \quad
m_{H^\pm} = 300\, \hbox{GeV}, \quad
m^{\mathrm{soft}}_{\tilde\ell} = 150\, \hbox{GeV}
 \; .
\label{point3l}
\end{eqnarray}
At this point, relevant masses and BR's are:
\begin{eqnarray}
m_{\widetilde\chi^\pm_{1,2}} = 116.85,~231.48\, \hbox{GeV},\quad & &
m_{\widetilde\chi^0_{1-4}}
= 87.93,~122.01,~140.29,~230.76\, \hbox{GeV},
\nonumber \\
{\mathrm{BR}}(H^- \rightarrow \widetilde\chi_1^-
\widetilde\chi_{2(3)}^0) =
0.18(0.03) \, ,\quad & &
{\mathrm{BR}}(H^- \rightarrow b\bar t) = 0.63 \, , 
\nonumber \\
{\mathrm{BR}}(\widetilde\chi_{2(3)}^0 \rightarrow \widetilde\chi_1^0
\ell^+
\ell^-) = 0.33(0.02) \, ,\quad & &
{\mathrm{BR}}(\widetilde\chi_1^- \rightarrow \widetilde\chi_1^0 \ell^-
\bar{\nu}_{\scriptscriptstyle \ell}) = 0.24 \; .
\label{p3lchar}
\end{eqnarray}
Additional key variables to be aware of are  
$m_{\widetilde\chi^-_{1}}-m_{\widetilde\chi^0_{1}}$ and
$m_{\widetilde\chi^0_{2,3}}-m_{\widetilde\chi^0_{1}}$.
These are not so large here and this softens both the
lepton spectra (the leptons coming from chargino and
neutralino decays have on average lower transverse
momenta than those coming from gauge boson decays)
and that of ${p}^{\mathrm{miss}}_T$.
Another point worthy of mention is that
the MSSM parameter point (\ref{point3l}),
gives $m_h = 105.5\, \hbox{GeV}$ {\em if} soft stop masses
are set to $1\, \hbox{TeV}$ and $A_t = 0$.  With a
projected LEP2 reach of $E_{cm} \simeq 208\, \hbox{GeV}$,   
this would yield a Higgsstrahlung cross-section 
that should be observable.
However, if, for instance, $A_t$ is raised to
$2\, \hbox{TeV}$, then $m_h = 118.9\, \hbox{GeV}$ and
on-shell Higgsstrahlung is kinematically forbidden.   
While soft SUSY-breaking parameters such as $A_t$ can
have a strong impact on $m_h$
(as well as a possibly significant impact on the 
$b \rightarrow s \gamma$ rates, as will be discussed later), 
they have very little effect on $m_{H^{\pm}}$ (as noted earlier).
Thus some care must be taken that all relevant free 
parameters in the model are adequately explored so as to 
not neglect allowable MSSM parameter sets.

The dominant SM backgrounds are again those involving
double- and single-top production and decay, 
this time accompanied by an additional
lepton-antilepton pair (electrons or muons)
produced in the `off-shell decay' of a neutral 
gauge boson ($V=\gamma/Z$)\footnote{To assess the $ttV^*$
background, the code originally developed in
\cite{ttVpap} was adapted to allow for an off-shell
gauge boson.}:
\begin{eqnarray}
gg, q\bar q\rightarrow t \bar t V^* \; ,  \quad
& t \rightarrow b q \bar q' \; , & \quad
\bar t \rightarrow \bar b \ell^- 
\bar{\nu}_{\scriptscriptstyle \ell} \; ,\quad
  V^* \rightarrow \ell^-\ell^+ \; ,
\label{ttV}
\\
\hbox{and} \qquad
gg,q\bar q \rightarrow t \bar b W^- V^* \; ,  \quad
& t \rightarrow b q \bar q' \; , & \quad
W^- \rightarrow \ell^- 
\bar{\nu}_{\scriptscriptstyle \ell} \; , \quad
V^* \rightarrow \ell^-\ell^+ \; .
\label{tV}
\end{eqnarray}

The set of cuts applied is similar that employed for the
one-lepton signal analysis:
\begin{enumerate}
\item 
Jets and leptons are retained if they satisfy the
following requirements:
$p_T^{\ell} >10\, \hbox{GeV}$,
$p_T^j >25\, \hbox{GeV}$, $|\eta_{\ell,j}|<2$ and
$\Delta R_{\ell,j/j,j}  > 0.4$.  The signal rate is 
sensitive to the $p_T$ threshold for the leptons ---
lowering the threshold will enhance the signal 
survival rate more than that of the backgrounds.
The value chosen here is reflective of the 
capabilities of the ATLAS detector \cite{ATLAS-TDR}.
As with the cuts for the one-lepton signal, for an 
event to pass this first cut it is compelled to have
no more than four jets fulfilling the requirements.
And in this case we of course demand that exactly
three leptons also satisfy the criteria.

\item 
We require that 
${p}^{\mathrm{miss}}_T > 25\, \hbox{GeV}$.

\item 
As before,
we impose $|M_{q\bar q'} - M_{W^\pm}|< 10\, \hbox{GeV}$.

\item 
We also again impose
$|M_{b q \bar q'} - m_t| < 25\, \hbox{GeV}$.

\item 
Given the rather low missing momenta involved in both signal
and backgrounds, we find that the variable
$\frac{|{p}^{\mathrm{miss}}_T-p_T^\ell|}   
      {|{p}^{\mathrm{miss}}_T+p_T^\ell|}$
used in the one-lepton analysis is no longer a suitable 
discriminant and do not include it in the cuts. 

\item 
Lastly, we apply a $Z^0$-veto,
$|M_{\ell^-\ell^+} - M_Z| > 10\, \hbox{GeV}$.
This is to eliminate the SM backgrounds where the gauge boson 
is on- or nearly on-shell.  Note that, for the signal,
$m_{\widetilde\chi^0_{2,3}}-m_{\widetilde\chi^0_{1}}\ll M_Z$,
and thus few signal events are lost here.
\end{enumerate}

Results for this series of cuts are given in Tab.$\!$~\ref{tab_3l}.
Unlike in the case of the one-lepton signature,
after cuts the three-lepton signal rate can be made
competitive with the background rates.  However, the total
number of signal events is low. If one multiplies the final row
of numbers of Tab.$\!$~\ref{tab_3l} by the 
$b$-tagging$\times$mis-tagging factor of
$\frac{1}{2} \! \times \! 0.96$ and by the leptonic $W^{\pm}$ and $Z^0$
BR's (for first two columns) or the leptonic branching ratios from 
Eqs.$\!$ (\ref{p3lchar}) (for last two columns), one finds for one year's
running ($100\, \hbox{fb}^{-1}$) the ratio,
$\hbox{\sl signal events} : \hbox{\sl background events} = 7:20$.
Note that the enhanced leptonic BR of $\widetilde\chi^0_{2}$
due to the light slepton intermediate state is very significant.

A more stringent cut can be applied on the invariant mass of the 
opposite-sign lepton pair if we take advantage of the fact that 
$M_{\ell^-\ell^+} < m_{\widetilde\chi^0_{2,3}}
\! -m_{\widetilde\chi^0_{1}} < M_Z$.  This entails the possession of some
information about the masses of the lowest-lying sparticle states.  As
noted earlier, it is quite likely that such information will be available
to those analyzing the real experimental data in search of a signal for 
the charged Higgs boson.  (Note that it should also be possible to tune
the $M_{\ell^-\ell^+}$ limit to help optimize any observed signal
even with incomplete information about the sparticle masses.)
Seeing that the $\chi^0_{1}\chi^0_{3}$ contribution is not very
important, we could impose the cut $M_{\ell^-\ell^+} < 
m_{\widetilde\chi^0_{2}}-m_{\widetilde\chi^0_{1}}$,
with the inclusion of which one finds for one year's running the ratio,
$\hbox{\sl signal events} : \hbox{\sl background events} = 5:5$.
Note that with the parameter set (\ref{point3l}) and the consequent 
sparticle mass spectrum (\ref{p3lchar}), cut 6. of 
Tab.$\!$~\ref{tab_3l} will be completely subsumed by this new cut.  
While this should be the case in general, it is safer to separately
apply cut 6. and this new cut since the choice of the former is not
parameter-space dependent and the cut-off for the later might rise 
above $M_Z - 10\, \hbox{GeV}$ in exceptional cases.

Another additional cut which has some dependence on the mass spectrum of
the charginos and neutralinos can be applied by defining 
$M_T(3\ell)\, \equiv \, \sqrt{2 p^{3\ell}_T p_T^{\mathrm{miss}}
(1-\cos\!\Delta\phi)}$,
where $p^{3\ell}_T$ is the transverse momentum of the three-lepton system
and $\Delta\phi$ is the azimuthal separation between
$p^{3\ell}_T$ and $p_T^{\mathrm{miss}}$.
Fig.$\!$ 5 illustrates how well this variable distinguishes our
signal from the backgrounds.  For the former, the $M_T(3\ell)$
distribution dies at $m_{H^\pm}-2m_{\widetilde\chi^0_1}$,
which is $\approx\! 123\, \hbox{GeV}$ for the point (\ref{point3l});
whereas, for the latter, it can stretch far beyond this value.
Demanding $M_T(3\ell) < 100\, \hbox{GeV}$ yields the ratio,
$\hbox{\sl signal events} : \hbox{\sl background events} = 7:9$ $(5:2)$,
if applied on top of the cuts in Tab.$\!$~\ref{tab_3l}
(or in conjunction with strengthening the 
$|M_{\ell^-\ell^+} - M_Z| > 10\, \hbox{GeV}$ cut in Tab.$\!$~\ref{tab_3l}
to $M_{\ell^-\ell^+} <
m_{\widetilde\chi^0_{2}}-m_{\widetilde\chi^0_{1}}$).

\begin{table}[!t]
\begin{center}
\begin{tabular}{|c|c|c||c||c|}
\hline
&&&&\\[-4mm]
& $t \bar{t} V^*$
& $t \bar{b} W^- V^*$
& $H^- \rightarrow \widetilde\chi^-_1 \widetilde\chi_2^0$
& $H^- \rightarrow \widetilde\chi^-_1 \widetilde\chi_3^0$\\
\hline
&      &      &      &     \\[-4mm]
No cuts
& 698. & 111. & 43.  & 7.0 \\
&      &      &      &     \\[-4mm]
\hline
&      &      &      &     \\[-4mm]
3 leptons each with  
$p_T^{\ell} >10\, \hbox{GeV}$,
&      &      &      &     \\
$|\eta_\ell| < 2$, $\Delta R(\ell,j)>0.4 $
& 317. & 49.7 & 8.7  & 2.2 \\
&      &      &      &     \\[-4mm]
\hline
&      &      &      &     \\[-4mm]
$\le 4$ jets with $p_T^j >25\, \hbox{GeV}$,
&      &      &      &     \\
$|\eta_j|<2$, $\Delta R(j,j)>0.4 $
& 161. & 30.3 & 1.75 & .43 \\ 
&      &      &      &     \\[-4mm]
\hline
&      &      &      &     \\[-4mm]
${p}^{\mathrm{miss}}_T > 25\, \hbox{GeV}$
& 133. & 21.1 & 1.67 & .42 \\
&      &      &      &     \\[-4mm]
\hline
&      &      &      &     \\[-4mm]
$|M_{q\bar q'}-M_{W^\pm}| <10\, \hbox{GeV}$
& 126. & 20.9 & 1.46 & .39 \\ 
&      &      &      &     \\[-4mm]
\hline
&      &      &      &     \\[-4mm]
$|M_{bq\bar q'}-m_t| < 25\, \hbox{GeV}$  
& 110. & 11.3 & 1.41 & .33 \\
\hline
&      &      &      &     \\[-4mm]
$|M_{\ell^-\ell^+}-M_Z| > 10\, \hbox{GeV}$
&  17. & 5.03 & 1.38 & .32 \\
\hline
\end{tabular}
\end{center}
\caption{\small
Production and decay rates (in femtobarns) for three-lepton signal 
and backgrounds, after the implementation of successive cuts.
Rates already include the Higgs boson decay BR's, whereas
a common hadronic BR for the decaying top ($0.699$) and a
common $b$-tagging$\times$mis-tagging factor of
$\frac{1}{2} \! \times \! 0.96$ have been omitted.
The SM backgrounds should also be multiplied by $2\times 0.212 \times 0.066$
(accounting for the $W^{\pm}$ and $Z^0$ leptonic BR's), while the signal
rates should be multiplied by
$2\times{\mathrm{BR}}(\widetilde\chi_1^- \rightarrow \widetilde\chi_1^0
\ell^-
\bar{\nu}_{\scriptscriptstyle \ell})
\times{\mathrm{BR}}(\widetilde\chi^0_{2(3)} \rightarrow \widetilde\chi_1^0  
\ell^- \ell^+)$.
The $t\bar t V^*$ and $t\bar b W^-V^*$ cross-sections ($V=\gamma,Z$) are
expressible in terms of the $Z^0$ decay rates since, at the end of the   
series of cuts, the $Z^0 \rightarrow \ell^- \ell^+$ contribution is
numerically dominant over the one from $\gamma^* \rightarrow
\ell^-\ell^+$.
Also note that, since $m_{\ell}$ is set to zero, a hard cut of
$M_{\ell^-\ell^+} >10\, \hbox{GeV}$ is necessary to avoid the
$\gamma^* \rightarrow \ell^-\ell^+$ singularity --- this is included
in the ``No cuts'' rates for $t\bar{b}W^- V^*$ and $t\bar{t}V^*$.
}
\label{tab_3l}
\end{table}

Bolstered somewhat by this result, it is reasonable to 
consider still higher charged Higgs boson masses.
Now competing factors come into play.  
One the one hand, the three-lepton chargino-neutralino 
decay channels remain large --- with 
$m_{H^\pm}=400\, \hbox{GeV}$ and all other MSSM parameters 
the same as in point (\ref{point3l}), one has
${\mathrm{BR}}(H^- \rightarrow \widetilde\chi_1^-\widetilde\chi_2^0)=0.14$
and
${\mathrm{BR}}(H^- \rightarrow \widetilde\chi_1^-\widetilde\chi_3^0)=0.03$
---
and, on the positive side, the lepton spectrum becomes harder.
On the other hand, the top-$H^{\pm}$ production rate drops 
precipitously, by more than a factor of 2 when $m_{H^\pm}$ is 
increased from $300\, \hbox{GeV}$ to $400\, \hbox{GeV}$.
Unfortunately the latter negative effect dominates: although
a greater percentage of the signal events survive
the cuts, one starts with a production cross-section that is 
just too small.
Applying the same cuts as in Tab.$\!$~\ref{tab_3l}
for $m_{H^{\pm}}=400\, \hbox{GeV}$ yields 
$0.55\, \hbox{fb}$ and $0.20\, \hbox{fb}$ 
for the $H^- \rightarrow \widetilde\chi_1^-\widetilde\chi_2^0$ and
$H^- \rightarrow \widetilde\chi_1^-\widetilde\chi_3^0$ channels,
respectively.
Adopting the same $b$-tagging efficiencies and 
leptonic BR's as for the $m_{H^{\pm}}=300\, \hbox{GeV}$
case now yields (for $m_{H^{\pm}}=400\, \hbox{GeV}$) the ratio, 
$\hbox{\sl signal events} : \hbox{\sl background events} = 3:20$.
In addition, the larger $m_{H^{\pm}}$ means the extra 
$M_T(3\ell)$ cut must be weakened --- requiring
$M_T(3\ell) < 120\, \hbox{GeV}$ 
along with $|M_{\ell^-\ell^+} - M_Z| > 10\, \hbox{GeV}$ leads to the
ratio,
$\hbox{\sl signal events} : \hbox{\sl background events} = 3:11$.
Also, since the $H^- \rightarrow \widetilde\chi_1^-\widetilde\chi_3^0$
decay modes acconts from more of the signal now, if the 
$M_{\ell^-\ell^+} < m_{\widetilde\chi^0_{2}}-m_{\widetilde\chi^0_{1}}$
cut is implemented, more signal will be lost, while a weaker
$M_{\ell^-\ell^+} < m_{\widetilde\chi^0_{3}}-m_{\widetilde\chi^0_{1}}$
cut is much less effective at cutting away backgrounds.
Thus, both the total three-lepton signal event rate and 
its statistical significance decline as $m_{H^{\pm}}$ is
raised from $300\, \hbox{GeV}$ to $400\, \hbox{GeV}$.
To this though must be added the caveat that, 
at $m_{H^{\pm}} = 400\, \hbox{GeV}$, decay modes 
including either the heaviest chargino or the heaviest
neutralino are significant (as seen from Fig.$\!$ 2).
We have neglected these, and so our results may be 
viewed as conservative. But it is nonetheless evident 
that $m_{H^\pm}=400$ GeV is near the kinematical limit
beyond which there is too little top-$H^{\pm}$ production 
cross-section at the LHC to exploit through the 
chargino-neutralino decay channels.  

In contrast, for $m_{H^{\pm}} \, \lsim \, 400\, \hbox{GeV}$, 
$\tan\!\beta$ values up to $\sim 10$ can be scanned
(for a significant portion of the possible values
of the other MSSM input parameters) using the trilepton 
plus top signature from
$H^{\pm} \rightarrow \widetilde\chi^{\pm}_1 \widetilde\chi^0_{2,3}$
decays.
In fact, over the range $3 \,\, \lsim \, \tan\!\beta \,\, \lsim \, 10$,
as $\tan\!\beta$ gets larger, the enhancement (or suppression)
of the Higgs decay rates is compensated by an opposite effect 
in the production rate.
This is understandable since the BR's for 
$H^{\pm} \rightarrow \widetilde\chi^{\pm}_i \widetilde\chi^0_j$
strengthen as the $H^- \rightarrow b\bar{t}$ decay width weakens
(as can be seen from an examination of Fig.$\!$ 2),
and the latter is proportional to the same coupling (\ref{Htbcoup}) 
as the top-$H^{\pm}$ production modes.  
Beyond $\tan\!\beta \approx 10$, the 
BR($H^{\pm} \rightarrow \widetilde\chi^{\pm}_i \widetilde\chi^0_j$)'s
start falling below the $\tan\!\beta = 4$ values simulated in 
the preceding numerical analyses.  In fact, 
due to the strengthening of other alternative decay modes
(such as $H^- \rightarrow \tau^- \bar{\nu}_{\tau}$), these BR's 
fall even quicker than the top-$H^{\pm}$ production rate increases.

Since the event rate imposes a discovery limit 
($\lsim \, 400\, \hbox{GeV}$)
on the mass of the charged Higgs bosons, it is important to determine if
all or part of the region of parameter space where the signal seems
observable is excluded by constraints coming from
$b \rightarrow s \gamma$ decays.  At present however, we are unaware
of any clean answer to this question.  To circumvent the
$b \rightarrow s \gamma$ constraint with such modest charged Higgs
boson masses, there must be sufficient cancellation between the
contributions from the top-charged Higgs boson loop and the chargino-stop
loop (we assume contributions from gluino-squark loops \cite{gluinoloop}
are negligible, though this need not always be true).  Seeing that our 
signal stems from charged Higgs boson decays including a chargino (this
prefers, among other things, lower values for $| \mu |$), it follows
that the main freedom left to adjust the
$b \rightarrow s \gamma$ rates lies in the stop sector.\footnote{
Here we are operating under the assumption that 
inter-generational squark mixing is negligible, analogous to the case 
for the third generation of the CKM matrix.  If squark mixings are 
made arbitrary, the $b \rightarrow s \gamma$ constraint can almost
certainly be evaded; however, constraints from other flavor-changing
neutral current processes need to be considered as well.}
The MSSM stop input parameters must be large enough to boost the light
Higgs boson mass above the LEP2 bounds (for the sample point we have 
used for illustration, this would translate into stop masses in the
${\sim}400$-$600\, \hbox{GeV}$ ballpark), yet low enough to ensure
sufficient cancellation among the MSSM $b \rightarrow s \gamma$
diagrams (see \cite{FCNCpap}, Fig.$\!$ 19, for a rough idea).
Studies addressing this have been done in the more restrictive
mSUGRA scenario (see for example Goto and Okada of \cite{bsgam};
note though that this is a leading-order calculation, and that more
recent next-to-leading order studies suggest there may be significant
modifications to these results --- see Ciuchini {\it et al.} and
Bobeth {\it et al.} of \cite{bsgam}), and point toward a severe
curtailment of the range of open parameter space for our signal.  
However, in mSUGRA the stop parameters are related to those of the 
Higgs bosons and the Higgsinos.  
If a more general MSSM scenario is assumed these relationships will
disappear and the restriction from $b \rightarrow s \gamma$ 
may well be significantly relaxed.

One final issue to consider is the distinguishability of
the charged Higgs bosons from the neutral $H$ and $A$.  
Throughout much of the parameter space, these MSSM Higgs 
bosons all have very similar masses.  Background to 
the charged Higgs boson signal could come from
$gg,q\bar{q} \rightarrow t\bar{t}H,t\bar{t}A$ 
%%production 
\cite{KunZwir,ttH0} or $qq' \rightarrow t\bar{b}H,t\bar{b}A$ production
\cite{tH0}, where the neutral MSSM Higgs boson then decays into 
chargino or neutralino pairs.  Fortunately, for the moderate 
$\tan\!\beta$ values we are interested in here, the $t\bar{t}H$ and
$t\bar{t}A$ production rates at the LHC are about an order of 
magnitude lower than the top-$H^{\pm}$ production rate. 
(See also paper \#2 of \cite{Whdec}.)

In summary, chargino-neutralino decays of 
heavy ({\it i.e.}, with $m_{H^\pm} \, \gsim \, m_t$) MSSM charged 
Higgs bosons have some potential to aid in the detection of 
such $H^{\pm}$ scalars at the LHC, especially in the intermediate 
$\tan\!\beta$ region, $3 \,\, \lsim \, \tan\!\beta \,\, \lsim \, 10$, 
which is inaccessible 
via SM decays.\footnote{Chargino-neutralino decays could of course 
also hinder detection in other $\tan\!\beta$ regions by suppressing  
the BR's of SM decay modes that do yield signals.  This also 
merits further attention.}  Among the possible 
chargino-neutralino combinations, the most promising are the
$H^- \rightarrow
\widetilde\chi_1^-\widetilde\chi_2^0,
\widetilde\chi_1^-\widetilde\chi_3^0$ 
decays followed in turn by the decays
$\widetilde\chi_1^- \rightarrow \widetilde\chi_1^0 \ell^- 
\bar{\nu}_{\scriptscriptstyle \ell}$
and 
$\widetilde\chi_{2,3}^0 \rightarrow \widetilde\chi_1^0\ell^+\ell^-$,
yielding a three-lepton final state.  
(The decay $H^- \rightarrow \widetilde\chi_1^-\widetilde\chi_1^0$
leading to a one-lepton final state was also studied, but found to 
be overwhelmed by the SM backgrounds.)
Since the charged Higgs boson is dominantly produced in 
association with a top quark, the main signature to look 
for is three hard, isolated leptons plus significant
${p}^{\mathrm{miss}}_T$ and a reconstructed top quark.  
Unfortunately, after the cuts utilized in this simulation,
the surviving signal event rates are small --- only around a 
handful of events per year.  Furthermore, these rates are sensitive to 
several MSSM parameters, including $M_{\scriptscriptstyle 2}$, 
$\mu$, and (via the leptonic BR's of $\widetilde\chi_{2,3}^0$)
$m^{\rm{soft}}_{\tilde{\ell}}$.  However, given the paucity 
of handles which can be used to study charged Higgs boson
production at the LHC for intermediate $\tan\!\beta$ values, 
further investigation
of even such weak signals in the more realistic environment 
of a full event generator, including parton shower effects
and hadronization, would be clarifying and beneficial.
Such a study is well underway (using the HERWIG \cite{herwig}
and ISAJET \cite{isajet} event generators) and we plan to 
report the results from these studies in the near 
future \cite{preparation}.

\vskip0.25cm\noindent
\underbar{\sl Acknowledgements}~
MB is grateful to the  U.S. National Science Foundation for
support under grant INT-9804704 and also to the
DESY theory group and the Physics Department of the Technion
for warm hospitality during portions of this study.
He also thanks P. Mercadante for very helpful correspondence.
MG is thankful to D. Zeppenfeld for fruitful discussions
and to the Theory Group at Saha Institute of Nuclear Physics
for hospitality during the final phase of this work.
SM is grateful to PPARC for financial support.
He also thanks K.A. Assamagan and Y. Coadou for useful conversations.
The authors thank F. Borzumati, M. Misiak and Y. Okada for 
helpful information concerning $b \rightarrow s \gamma$.
They also thank X. Tata and P. Zerwas for their careful reading of 
the manuscript and their sagacious suggestions.

\def\baselinestretch{1.0}
\vfill\clearpage\thispagestyle{empty}
\noindent
{Figure 1: BR's of the MSSM charged Higgs boson into
chargino-neutralino pairs (summing all such channels),
in the $(m_A,\tan\beta)$ plane,
with $M_{\scriptscriptstyle 2} = 200\, \hbox{GeV}$ 
and $\mu = -120\, \hbox{GeV}$.
One-loop formul\ae\ as found in \cite{thesis,isajet}
are used to relate $m_{H^{\pm}}$ to $m_A$.
Other MSSM input parameters are:
$m^{\rm{soft}}_{\tilde{q}} = 1\, \hbox{TeV}$, 
$A_t = 2\, \hbox{TeV}$ and
$m^{\rm{soft}}_{\tilde{\ell}} = 300\, \hbox{GeV}$.}
\includegraphics{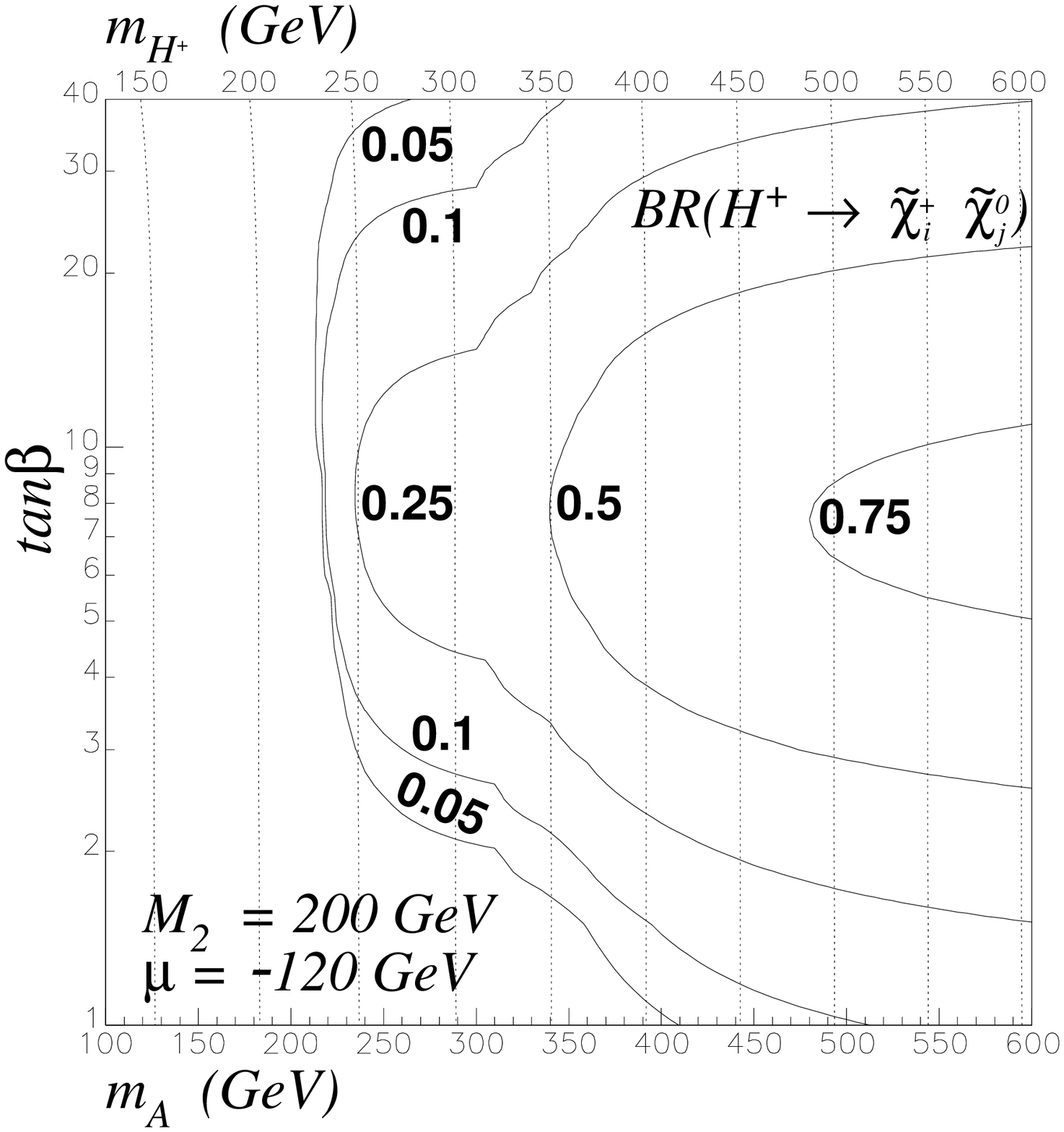}

\vfill\clearpage\thispagestyle{empty}
\noindent
{Figure 2: BR's of the MSSM charged Higgs boson 
as a function of $\tan\!\beta$ for 
$m_{H^\pm} = 200$, $300$, and $400\, \hbox{GeV}$,
again with $M_{\scriptscriptstyle 2} = 200\, \hbox{GeV}$ and
$\mu = -120\, \hbox{GeV}$. Other MSSM input parameters 
are also as in Fig.$\!$~1, except that here
$m^{\rm{soft}}_{\tilde{\ell}} = 150\, \hbox{GeV}$.}
\includegraphics{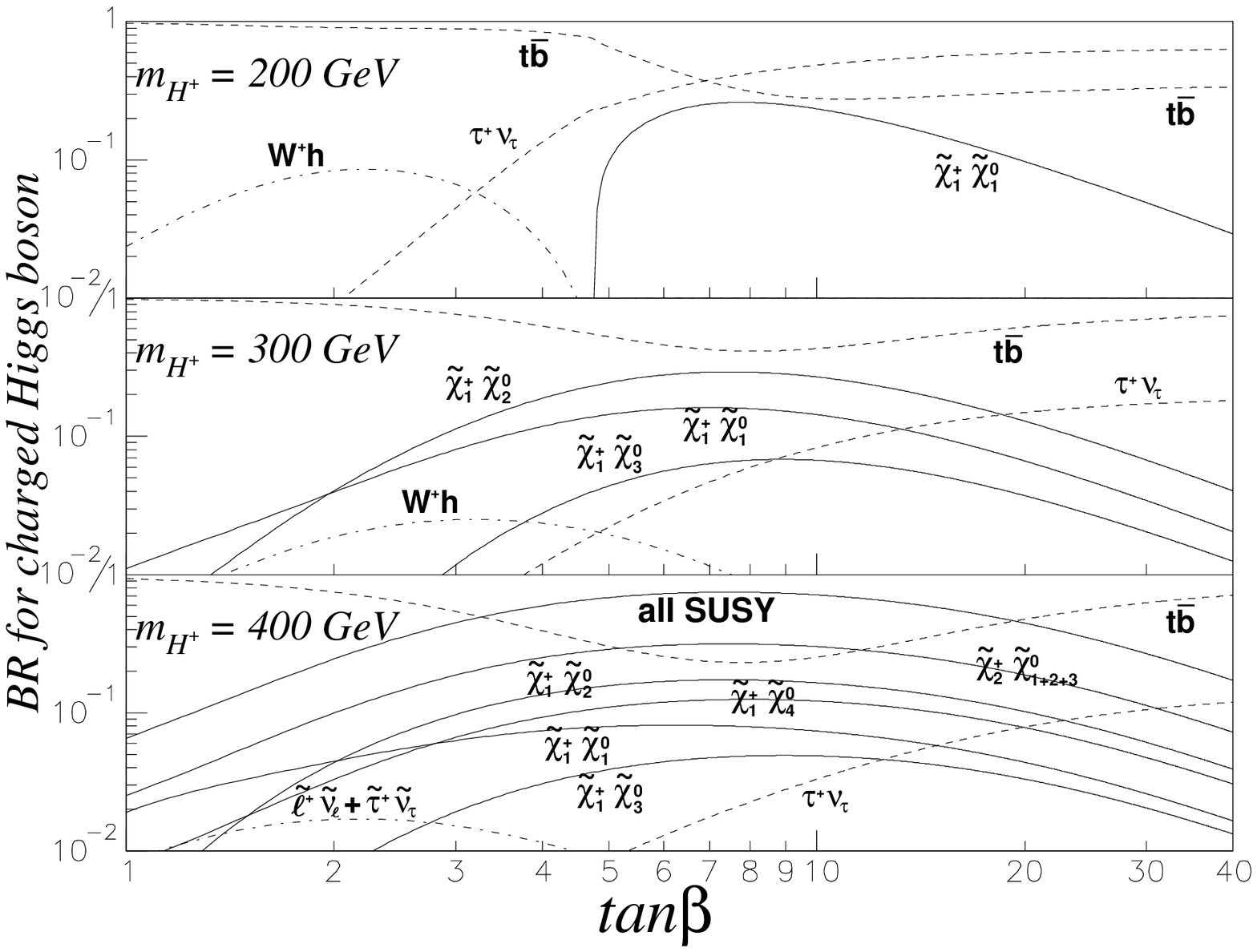}

\vfill\clearpage\thispagestyle{empty}
\noindent
{Figure 3: BR($H^+ \rightarrow \widetilde\chi_1^+ \widetilde\chi_1^0$)
in the $(\mu,M_{\scriptscriptstyle 2})$ plane 
for $\tan\!\beta=4$ and $m_{H^{\pm}} = 200\, \hbox{GeV}$. 
Other MSSM input parameters are as in Fig.$\!$~1.
The dotted lines for $m_{{\chi}^{\pm}_1}$ indicate the reach
of LEP2.  As expected, setting
$m^{\rm{soft}}_{\widetilde{\ell}} = 150\, \hbox{GeV}$ or
$m^{\rm{soft}}_{\widetilde{\ell}} = 300\, \hbox{GeV}$ does not 
affect the results.
}
\includegraphics{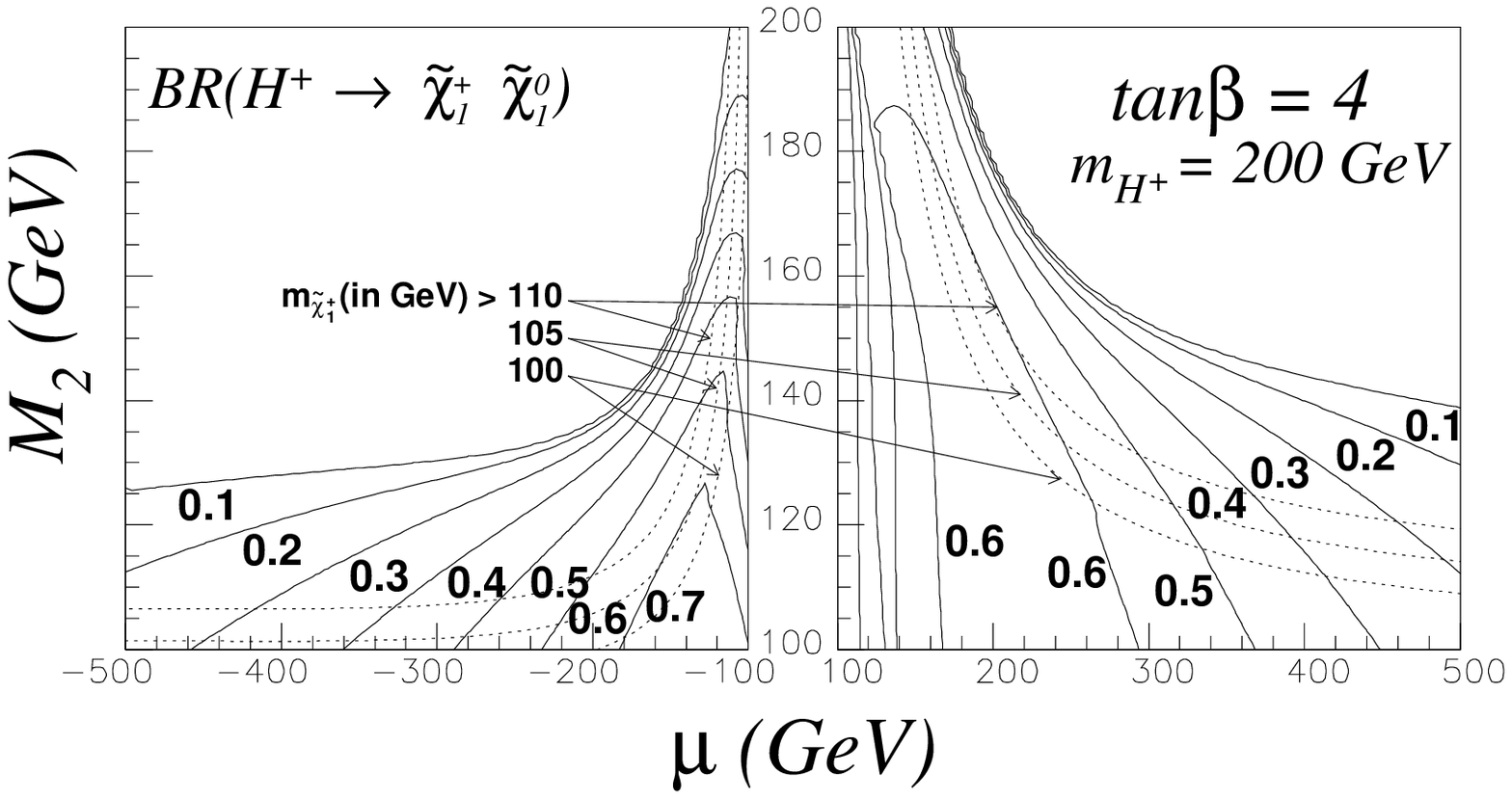}

\vfill\clearpage\thispagestyle{empty}
\noindent
{Figure 4: BR($H^+ \rightarrow \widetilde\chi^+_1 \widetilde\chi_2^0$)
(on top) and BR($H^+ \rightarrow \widetilde\chi^+_1 \widetilde\chi_3^0$)
(on bottom) in the $(\mu,M_{\scriptscriptstyle 2})$ plane  
for $\tan\!\beta=4$ and $m_{H^{\pm}} = 300\, \hbox{GeV}$.
Other MSSM input parameters are as in Fig.$\!$~1,
except that $m^{\rm{soft}}_{\tilde{\ell}} = 150\, \hbox{GeV}$.
The dotted lines for $m_{{\chi}^{\pm}_1}$ indicate the reach of 
LEP2.} 
\includegraphics{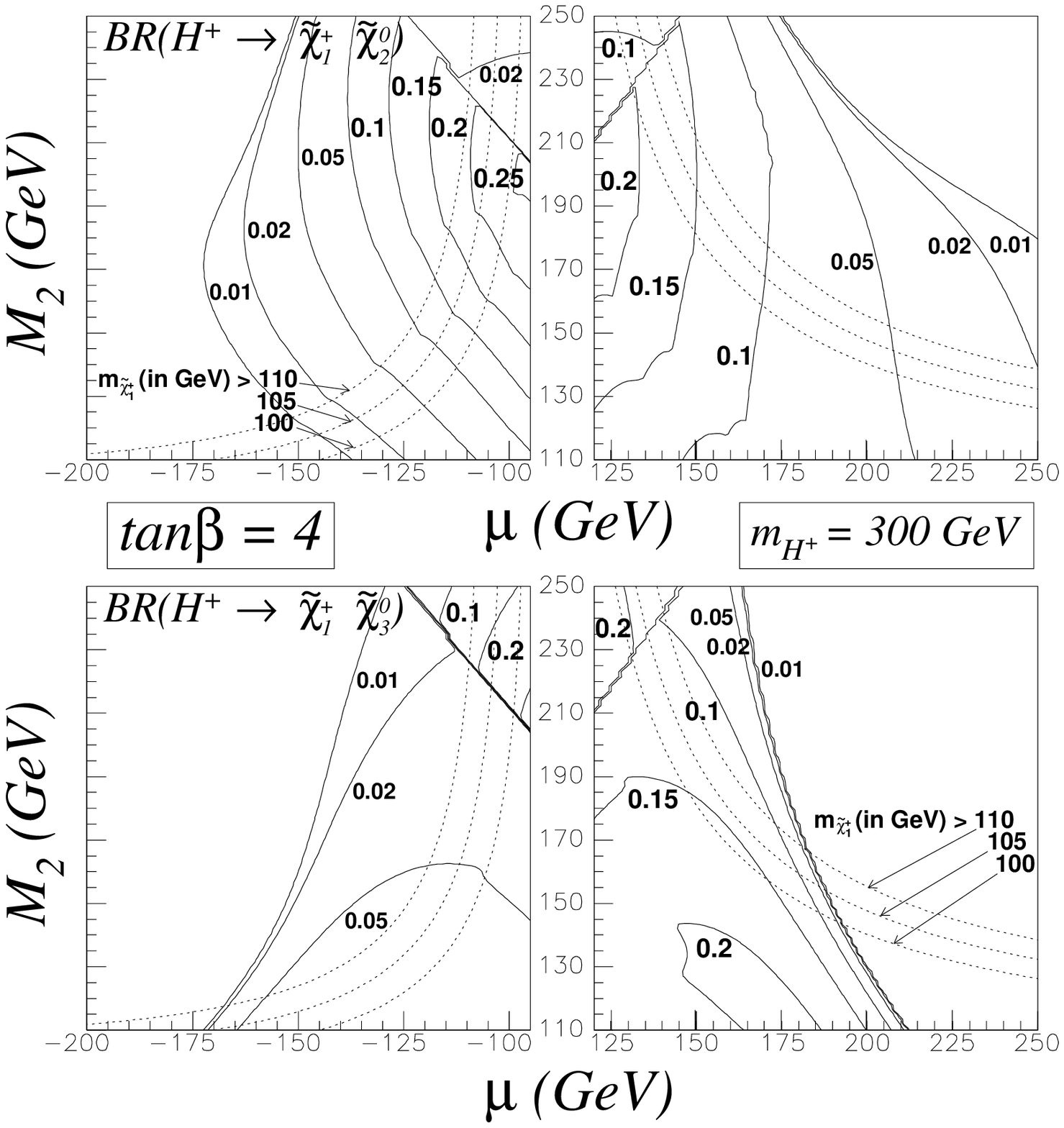}

\vfill\clearpage\thispagestyle{empty}
\noindent
{Figure 5: Normalized differential distributions 
 of the three-lepton system transverse mass, $M_T(3\ell)$ 
 (as defined in the text) 
 for: $H^+ \rightarrow \widetilde\chi^+_1 \widetilde\chi_2^0$
(solid: $m_{H^{\pm}} = 300\, \hbox{GeV}$;
dashed: $m_{H^{\pm}} = 400\, \hbox{GeV}$),
 $H^+ \rightarrow \widetilde\chi^+_1 \widetilde\chi_3^0$
(fine-dotted: $m_{H^{\pm}} = 300\, \hbox{GeV}$;
  dot-dashed: $m_{H^{\pm}} = 400\, \hbox{GeV}$),
 $t \bar{t} V^*$ (long-dashed),
and $t \bar{b} W^- V^*$ (dotted).
 Here $M_{\scriptscriptstyle 2} = 200\, \hbox{GeV}$, 
$\mu = -120\, \hbox{GeV}$ and $\tan\!\beta=4$. 
Other MSSM input parameters are as in Fig.$\!$~1. 
Normalizations are as after the cuts in 
Tab.$\!$~\ref{tab_3l}; {\it i.e.},
the leptonic BR's and the $b$-tagging$\times$mis-tagging 
efficiency are not included.}
\vskip2.0cm
~\epsfig{file=chfig5.ps,height=14cm,angle=90}
\end{document}